# Spin-polarized currents generated by magnetic Fe atomic chains

Zheng-Zhe Lin[1]* and Xi Chen[2]

*1) School of Physics and Optoelectronic Engineering, Xidian University, Xi'an 710071, China*

*2) Department of Applied Physics, School of Science, Xi'an Jiaotong University, Xi'an 710049, China*

*Corresponding Author. E-mail address: linzhengzhe@hotmail.com

**Abstract** – Fe-based devices are widely used in spintronics because of high spin-polarization and magnetism. In this work, free-standing Fe atomic chains were proposed to be used as the thinnest wires to generate spin-polarized currents due to the spin-polarized energy bands. By *ab initio* calculations, the zigzag structure was found more stable than the wide-angle zigzag structure and has higher ratio of spin-up and spin-down currents. By our theoretical prediction, Fe atomic chains have sufficiently long thermal lifetime only at $T \leq 150$ K, while C atomic chains are very stable even at $T=1000$ K. This result means that the spintronic devices based on Fe chains could only work at low temperature. A system constructed by a short Fe chain sandwiched between two graphene electrodes was proposed as a spin-polarized current generator, while a C chain does not have such property. The present work may be instructive and meaningful to further practical applications based on recent technical development on the preparation of metal atomic chains [*Proc. Natl. Acad. Sci. U.S.A.* **107**, 9055 (2010)].

## I. Introduction

Spintronics is a promising field for information processing, storage and many other applications. In the past decade, much effort has been made in spintronics and great developments have been seen [1-3]. In this area, generation and manipulation of spin-polarized current are two basic points, and minimization of spintronic devices is increasingly important. Since the discovery of graphene, more and more applications based on it were proposed. Due to half-metallic nature for edge states in zigzag graphene nanoribbons [4, 5], i.e. metallic for electrons with one spin orientation and insulating for electrons with the other, graphene-based spin-filter devices were proposed. However, edges of graphene suffer from spontaneous reconstruction [6, 7] or chemical passivation which may change the half-metallic electronic property. On the other hand, as the thinnest molecular wires one-dimensional atomic chains have gained great attention for their possible spintronic applications. Over the past two decades, Au [8-11] and Pt [12, 13] atomic chains were prepared by pulling two contacted atom-sized metal junctions. In recent years, free-standing C chains were carved out from single-layer graphene by high-energy electron beam [14], unraveled from sharp C specimens [15-17] or C nanotubes [18] and chemical synthesized [19-21]. Recently, much attention was paid to metal atomic chains and their stability was studied [22, 23]. By elongating a C nanotube with Fe nanorod inside, Fe atomic chain clamped by C nanotube was prepared [24]. And chemical reactivity and sensitivity of Au and Ag chains to small molecules were considered [25-27]. Among these atomic chains, magnetic chains may be the thinnest material used for generating and transporting spin-polarized currents because of high spin-polarization and magnetism, i.e. Fe atomic chains [28]. In recent years, two-dimensional graphene was considered as a good candidate for materials used in emerging electronics. In C-based circuits, one-dimensional C chains are the thinnest natural wires, and Fe chain embedded in C chains may work with the circuits. C chains have been proved to play a good role as spin-filter and spin-valve [29, 30], and magnetic Fe chain embedded in C chains should have a nice performance on spin-polarized current generation. To design Fe-based low-dimensional spintronic devices, theoretical investigation on the thermal stability and electronic property of Fe chains should be beneficial to guild corresponding experiments.

In this work, the spintronic property of Fe atomic chains was studied. On this basis, a short Fe atomic chain clamped by graphene electrodes with C chains [Fig. 3(a)] was proposed as a spin-polarized current generator. To investigate the thermal stability of Fe atomic chains, a recent built statistical mechanical model [31-34] was employed. According to the result, the C-C and C-Fe bonds in atomic chains are very stable at room temperature, while Fe-Fe bonds could only survive below 150 K with a lifetime longer than 99 hours. A designed structure composed by graphene electrodes, C chains and a Fe chain presented a feature of spin-polarized quantum electronic transport. The structure could be used as a smallest and thinnest spin-polarized current generator.

## II. Method

To investigate the electronic properties of Fe atomic chains and corresponding structures, density functional theory calculations were performed using the SIESTA code [35]. The exchange-correlation functional was treated using a generalized gradient approximation according to Perdew-Burke-Ernzerhof [36]. The norm-conserving Troullier-Martins pseudopotentials [37] were used to describe the core electrons. For structure optimization, the double-ζ plus polarization basis sets were used and the grid mesh cutoff was set 150 Ry. The structures were relaxed until the atomic forces become less than 0.01 eV/Å. For molecular dynamics (MD) simulations, non self-consistent Harris functional was used to save computation time.

For quantum transport, calculations were performed using the TRANSIESTA module [38]. For a bias voltage $V_b$, the spin-up part $I_+$ and spin-down part $I_-$ of the current is given by Landauer-Buttiker formula [39]

$$I_{\pm} = \frac{2e}{h}\int T_{\pm}(E,V_b)[f_L(E-E_F-\frac{eV_b}{2})-f_R(E-E_F+\frac{eV_b}{2})]dE. \qquad (1)$$

Here, $T_+(E, V_b)$ and $T_-(E, V_b)$ are the transmission rate of spin-up and spin-down electrons at energy $E$, respectively. $E_F$ is the Fermi energy of electrodes. $f_L$ and $f_R$ are Fermi-Dirac distribution functions for both electrodes, respectively. To save computation time, the single-ζ plus polarization basis sets were used.

A simple statistical mechanical model [31-34] built previously was applied to predict the lifetime of chemical bonds in atomic chains at different temperatures. In atomic chains, an element process may involve a transfer of 'key atom' in a potential valley crossing over $E_0$. In most cases the atomic kinetic energy $\varepsilon$ at the valley bottom is significantly smaller than $E_0$, and the atom vibrates many times in the valley before crossing over the barrier. Based on the Boltzmann distribution, the probability $P$ of the atomic kinetic energy $\varepsilon$ larger than $E_0$ reads,

$$P = \frac{\int_{E_0}^{+\infty} \varepsilon^{1/2} e^{-\varepsilon/k_B T} d\varepsilon}{\int_0^{+\infty} \varepsilon^{1/2} e^{-\varepsilon/k_B T} d\varepsilon} = \frac{\int_{E_0}^{+\infty} \varepsilon^{1/2} e^{-\varepsilon/k_B T} d\varepsilon}{\sqrt{\pi}(k_B T)^{3/2}/2}, \quad (2)$$

where $\varepsilon^{1/2} e^{-\varepsilon/k_B T}$ the Boltzmann distribution and $\int_0^{+\infty} \varepsilon^{1/2} e^{-\varepsilon/k_B T} d\varepsilon$ the partition function. Then, the atomic transfer rate $k_{1st}$ over the barrier reads

$$k_{1st} = k_0 \frac{\int_{E_0}^{+\infty} \varepsilon^{1/2} e^{-\varepsilon/k_B T} d\varepsilon}{\sqrt{\pi}(k_B T)^{3/2}/2} \quad (3)$$

and the thermal lifetime of the chemical bond $\tau = 1/k_{1st}$. Here, $k_0$ is the attempt frequency of key atom vibrating in the potential valley, which can be evaluated by the potential energy $U=U(s)$ along the reaction path with $ds = \left(\sum_{i=1}^{n} m_i d\vec{r}_i^{\,2}\right)^{1/2}$ the reaction coordinate. The Lagrangian along the reaction path is

$$L = \frac{1}{2}\left(\frac{ds}{dt}\right)^2 - U, \quad (4)$$

and the corresponding Lagrange's equation approximately reads

$$\frac{d^2 s}{dt^2} + k_0^2 s = 0, \quad (5)$$

where $k_0 = \left.\frac{d^2 U}{ds^2}\right|_{s=0}$ is just the attempt frequency. In our previous work, this model has been widely verified by an amount of reaction rate data from MD simulations [31-34].

## III. Results and discussion

3.1 *General information of infinite Fe atomic chains*

To get basic information, geometry optimizations and energy band calculations were performed for infinite Fe atomic chains. Linear and zigzag Fe chains were taken into account as possible structures, and considering the Peierls distortion a unit cell with two atoms was used. According to the result, the zigzag structure (upper Fig. 1(a)) has the lowest potential energy, and with the Peierls distortion the corresponding bond lengths are $c_1$=2.37 Å, $c_2$=2.55 Å, $c$=2.62 Å. The cohesive energy and the magnetic moment per unit cell are $E_c$=2.64 eV/atom and $\mu$=3.80 $\mu_B$, respectively. The above result is close to Ref. [40] ($c_1$=2.24 Å, $c_2$=2.42 Å, $c$=2.40 Å, $E_c$=2.69 eV/atom and $\mu$=3.19 $\mu_B$, by ultrasoft pseudopotentials and plane-wave basic set). It is worth nothing that in some papers Fe chain was simply considered as linear atomic chain [28, 41, 42] which may be not the most stable structure. Fig. 1(b) (upper) shows the energy band profile and density of state (DOS) for the zigzag Fe atomic chain, presenting a spin-polarized feature. To investigate the quantum transport property, a model was built by one and six unit cells of zigzag Fe chain selected as the electrode and the scattering region, respectively (Fig. 1(c)). A 1×1×100 *k*-point sampling was used for the electrodes, and a vacuum layer of 20 Å was used along *x* and *y* directions in order to minimize the interaction between neighboring unit cells. The result (upper Fig. 1(c)) indicates that the current is spin-polarized and $I_+$ : $I_-$=3.2 at $V_b$=1.0 V. In the range of $V_b$=0.0~1.0 V, the average conductance for spin-up and spin-down electrons are about 1.11 and 0.35 $G_0$ ($G_0$=$2e^2/h$ is the quantum conductance), which is close to the values in Ref. [43].

In experimental manipulation of nanostructures such as mechanically controllable break junctions [8-13], an atomic chain can be elongated. To investigate the effect of elongation on Fe chain, we enlarged the lattice constant *c* of zigzag Fe chain step by step and performed geometry optimizations. According to the result, another local minima of potential energy was found at $c$= 4.68 Å, which is elongated from zigzag Fe chain and can be called as wide-angle zigzag Fe chain (lower Fig. 1(a)). This stable structure has a smaller cohesive energy $E_c$=1.77 eV/atom than the zigzag structure and a magnetic moment $\mu$=3.57 $\mu_B$ per unit cell. Fig. 1(b) (lower) shows the energy band profile and DOS of the wide-angle zigzag Fe atomic chain, presenting a spin-polarized feature. For quantum transport, the current is less spin-polarized than that of zigzag chain (lower Fig. 1(c)), with a ratio of $I_+$: $I_-$=1.2 at $V_b$=1.0 V. In the range of $V_b$=0.0~1.0 V, the average conductance for spin-up and spin-down electrons are about 0.49 and 0.42 $G_0$.

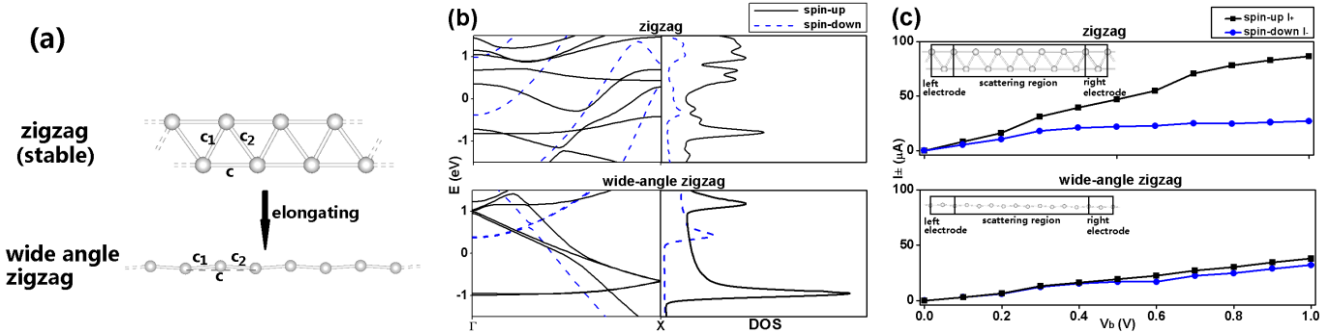

Fig. 1 (a) The zigzag structure of infinite Fe atomic chain (upper), and the wide-angle zigzag structure (lower) which is elongated from the zigzag one. (b) The energy bands and DOS of zigzag (upper) and wide-angle zigzag (lower) Fe atomic chains, where Γ the center and X the boundary of Brillouin zone. (c) The spin-up and spin-down current $I_+$ and $I_-$ at different bias $V_b$ for zigzag (upper) and wide-angle zigzag (lower) Fe atomic chains.

3.2 *Thermal lifetime of $C_8$, $C_4FeC_4$ and $C_4Fe_4C_4$ chain*

The stability of Fe atomic chains is vitally important in practical applications and must be investigated. Recently, a theoretical model was proposed to formulate criteria for the producibility of free-standing metal atomic chains in mechanically controllable break junctions [22], and the producibility of C atomic chains was also studied [44, 45]. Here, we concern thermal lifetime of Fe chains connected with C chains, i.e. $C_8$. $C_4FeC_4$ and $C_4Fe_4C_4$ chain (Fig. 2(a)). In thermal motion, a chemical bond in the chain may break when two neighboring atoms move away from each other (e.g. the progress shown in the panel of Fig. 2(a)), and average lifetime of the bond $\tau$=1/$k_{1st}$, where $k_{1st}$ can be evaluated by Eq. (3) and (5). To apply Eq. (3) and (5), potential profiles along minimum energy paths (MEPs) for bond breaking of C-C in $C_8$ chain, C-Fe in $C_4FeC_4$ chain and middle Fe-Fe in $C_4Fe_4C_4$ chain (Fig. 2(b)) were calculated with two terminal chain atoms fixed. As an example, the MEP for bond breaking of middle Fe-Fe bond in $C_4Fe_4C_4$ chain is shown in Fig. 1(a). For the above bond breaking progresses, we got $E_0$=5.44, 5.17 and 0.51 eV and $k_0$=8.05×10$^{11}$, 1.93×10$^{11}$ and 5.75×10$^{10}$ s$^{-1}$ (by Eq. (5)), respectively. Then by Eq. (3), average lifetime $\tau$ of the bonds at different temperatures was derived (Fig. 2(c)). According to Eq. (3) and (5), the middle C-C bond in $C_8$ chain has a lifetime of about 6×10$^{70}$ years at $T$=300 K. Even at $T$=1000 K, this bond still has a lifetime of about 1×10$^7$ years. For C-Fe bond in $C_4FeC_4$ chain, the lifetime is about 6×10$^{66}$ and 2×10$^6$ years at $T$=300 and 1000 K, respectively. The result indicates that the pure and Fe-doped C chains are very stable. However, at $T$=300 K the middle Fe-Fe bond in $C_4Fe_4C_4$ chain has a lifetime of only 1.30 ms. Fortunately, at $T\leq$150 K the lifetime is predicted to be longer than 99.0 hours. By *ab initio* MD simulation, we found that at $T$=300 K the middle Fe-Fe bond of $C_4Fe_4C_4$ chain kept stable even in a time of 50 ps, while survived for only 0.2 ps at $T$=1000 K. This result is generally in accordance with the above theoretical prediction. For this reason, Fe atomic chains may serve as spintronic devices at low temperature.

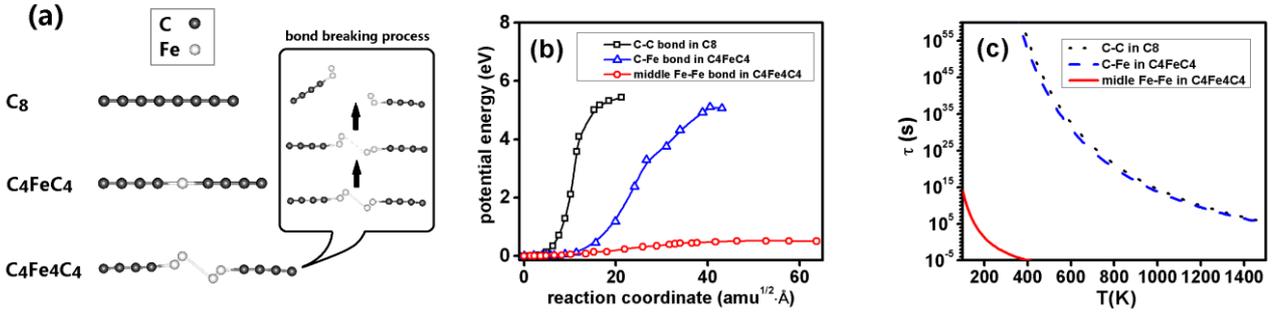

Fig. 2 (a) The structures of $C_8$, $C_4FeC_4$ and $C_4Fe_4C_4$ atomic chain. The right panel shows some snapshots along the MEP for bond breaking of the middle Fe-Fe bond. (b) Potential energy profile along the MEP for bond breaking of C-C in $C_8$ chain, C-Fe in $C_4FeC_4$ chain and middle Fe-Fe in $C_4Fe_4C_4$ chain. (c) The average lifetime $\tau$=1/$k_{1st}$ of these three bonds changing with temperature $T$.

3.3 *Spin-polarized electronic transport*

Based on optimized structures, we investigated the spintronic properties of Fe atomic chain coupled with C chain, which is the thinnest magnetic wire in graphene-based circuit. A two-probe system, with a $C_4Fe_4C_4$ atomic chain as the scattering region sandwiched between two semi-infinite graphene electrodes, was constructed along $z$ direction (Fig. 3(a)). Periodic boundary conditions are imposed on the electrodes (on *x* and *z* directions) and the scattering region (on *x* direction) as shown by the solid frame. Different lengths of buffer layers in the scattering region were tested to investigate the influence between the two electrodes, and three unit cells of the electrode were found to be enough. The zigzag direction of graphene electrodes was chosen as *z* direction, because the *k* vectors of the electronic states near the Fermi energy, i.e. on the Dirac point, are along *z* direction and thus the conductivity of the zigzag direction is better than the armchair direction [46]. A $4\times1\times100$ *k*-point sampling was used for the electrodes. The vacuum layer along *y* direction is 20 Å in order to minimize the interaction between neighboring sheets.

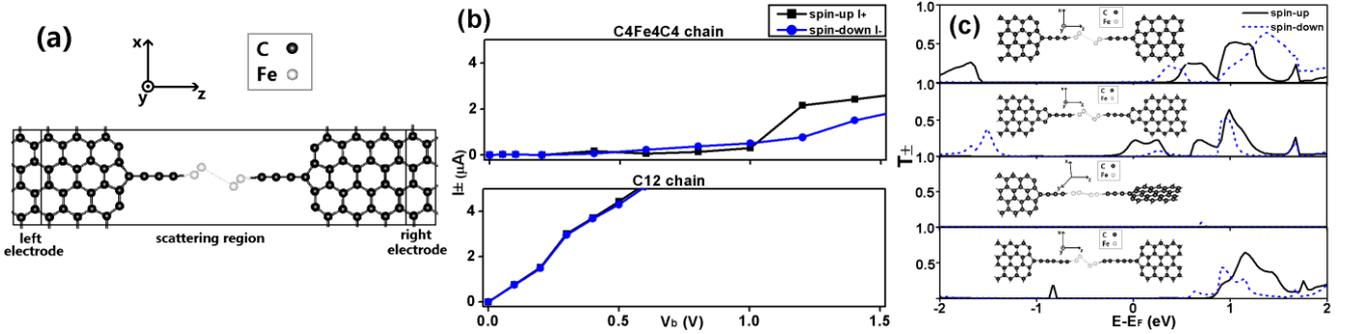

Fig. 3 (a) A two-probe system constructed by a $C_4Fe_4C_4$ atomic chain as the scattering region sandwiched between two semi-infinite graphene electrodes. Periodic boundary conditions are shown by the solid frame. (b) The spin-up and spin-down current $I_+$ and $I_-$ at different bias $V_b$ for a $C_4Fe_4C_4$ (upper) chain and a $C_{12}$ (lower) chain sandwiched between graphene electrodes. (c) Transmission spectra at $V_b=0$ V for four two-probe systems including C-Fe-C atomic chain. Corresponding scattering regions are shown in each subfigure. The solid and dashed lines indicate $T_+$ and $T_-$ for spin-up and spin-down electrons, respectively.

According to the result (upper Fig. 3(b)), for $V_b=0.6\sim1.0$ V, the spin-down current $I_-$ is larger than the spin-up current $I_+$, presenting a spin-polarized feature. The ratio $I_-$: $I_+$=3.8 at $V_b$=0.6 V, and decreases to $I_-$: $I_+$=1.7 at $V_b$=1.0 V. However, in this voltage range the current is quite small, and the average conductance for spin-up and spin-down electrons are only 0.008 and 0.009 $G_0$. For $V_b>1.0$ V, the current is larger than that for $V_b=0.6\sim1.0$ V, and the spin-up current $I_+$ is obviously larger than the spin-down current $I_-$. At $V_b$=1.4 V, the ratio $I_+$: $I_-$=1.6. By comparison, at the same $V_b$, $I_+$ and $I_-$ of this system are in an order of magnitude smaller than those of infinite Fe atomic chain. In the range of $V_b=1.0\sim1.4$ V, the average conductance for spin-up and spin-down electrons are about 0.05 and 0.03 $G_0$. To investigate the mechanism of spin-polarized currents, the transmission spectrum at $V_b=0$ V for the two-probe system in Fig. 3(a) is plotted in the first subfigure of Fig. 3(c). Transmission peaks for spin-up and spin-down electrons can be found at different energy $E$. These peaks should correspond to molecular electronic states of the scattering region. In the range of $E-E_F=0\sim1$ eV, a spin-down transmission peak locates at an energy range lower than a spin-up peak. Therefore, with increasing $V_b$ the spin-down current $I_-$ is larger and then smaller than the spin-up current $I_+$. In the range of $E-E_F>1$ eV, another spin-up transmission peak appears with higher transmission than the spin-down peak in this range, and correspondingly, for $V_b>1.0$ V $I_+$ is larger than $I_-$.

To investigate the influence of different shapes of electrode ends or lengths of C chain on the spin-polarized currents, three other structures (shown in the second to fourth subfigures of Fig. 3 (c)) were considered, which are the one with taper electrode ends, the one whose electrode planes are perpendicular to each other and the one with longer C chains, respectively. For the structure with taper electrode ends, the transmission of spin-up electrons is always larger than spin-down electrons in the range of $E-E_F=0\sim2$ eV, indicating that the currents are spin-polarized. For the structure whose electrode planes are perpendicular to each other, the transmission of spin-up and spin-down electrons is nearly zero in the whole energy range, i.e. the conductance is very low at any $V_b$. This is because in this case the $p$-$\pi$ conjugated orbitals in graphene electrodes are orthogonal to each side and then the transmission is forbidden. It can be inferred that when the two electrodes have an angle $\theta$ the transmission $T_\pm$ is proportional to $\cos^2\theta$. For the structure with longer C chains, the transmission of spin-up electrons is also larger than spin-down electrons in the range of $E-E_F>1$ eV. In summary, for $V_b>1$ eV, the spin-polarized currents by Fe atomic chain are independent of the geometry of electrodes or C chains but weaken by the increase of angle between the two electrodes.

Finally, to make a comparison with pure C atomic chain, we replaced four Fe atoms with four C atoms, and $I_+$ and $I_-$ of the optimized structure are shown in lower

Fig. 3(b). It can be seen that the pure C chain exhibits typical characteristic of spin-non-polarized conductor with a conductance of about 0.11 $G_0$ for spin-up and spin-down electrons. According to the above results, Fe atomic wire could play the role of spin-polarized current generator in C-based circuits.

## IV. Conclusion

Through the above theoretical investigation, free-standing Fe atomic chains were proposed to be used as the thinnest wire to generate spin-polarized currents. Long Fe chain has a stable zigzag structure and a spin-polarized energy band. At a bias of 1.0 V, the ratio of spin-up and spin-down current in zigzag Fe chain is about 3.2**:** 1. When zigzag Fe chain is elongated, it transforms into a wide-angle zigzag structure, whose ratio of spin-up and spin-down current is lower than that of zigzag structure (1.2**:** 1 at 1.0 V). By our theoretical prediction, Fe atomic chains have sufficiently long thermal lifetime only at $T \leq 150$ K, while C atomic chains are very stable even at $T=1000$ K. This result means that the spintronic devices based on Fe chains could only work at low temperature. A short Fe chain sandwiched between two graphene electrodes was taken as an example for spintronic devices, and the quantum electronic transport was investigated. At certain bias, the currents through this system are spin-polarized, while similar property did not found for pure C chains. By recent developed technique for preparing Fe chains clamped by C nanotubes [24], the spintronic property of Fe chains may be put into future practical applications.

∗∗∗

**Acknowledgements**

This work was supported by the National Natural Science Foundation of China under Grant No. 11304239, and the Fundamental Research Funds for the Central Universities.